\documentclass[epdftex,twocolumn,epjc3]{svjour3}

\RequirePackage[T1]{fontenc}

\smartqed  

\RequirePackage{graphicx}
\RequirePackage{mathptmx}      
\RequirePackage{flushend}
\RequirePackage[numbers,sort&compress]{natbib}
\RequirePackage[colorlinks,citecolor=blue,urlcolor=blue,linkcolor=blue]{hyperref}

\usepackage{graphics}
\usepackage{graphicx}
\usepackage{subfigure}
\usepackage{xcolor}
\usepackage{mathtext,bbm,amsmath,amsfonts,amssymb,indentfirst,syntonly,graphicx}
\usepackage{mathtools}
\usepackage{slashbox}
\usepackage[english]{babel}
\usepackage{calc}
\usepackage{tikz}
\journalname{Eur. Phys. J. C}

\begin{document}

\def\bc{\begin{center}}
\def\ec{\end{center}}
\def\be{\begin{eqnarray}}
\def\ee{\end{eqnarray}}

\title{Investigation of homogeneity and matter distribution \\on large scales using large quasar groups}
%
%

\author{Ming-Hua Li\thanksref{e1,addr1}
        \and
        Zhi-Bing Li\thanksref{e2} 
}

\thankstext{e1}{e-mail: limh@ihep.ac.cn}
\thankstext{e2}{e-mail: stslzb@mail.sysu.edu.cn}

\institute{School of Physics and Engineering, Sun Yat-Sen University, Guangzhou 510275, China\label{addr1}
}

\date{Received: date / Revised version: date}
%

\maketitle
\abstract{
We use 12 large quasar group (LQG) samples to investigate the homogeneity of $0.5\lesssim z \lesssim 2$ Universe ($z$ denotes the redshift). We calculate the bias factor $b$ and the two-point correlation function $\xi_{{\rm LQG}}$ for such groups for three different density profiles of the LQG dark matter halos, i.e. the isothermal profile, the Navarro-Frenk-White (NFW) profile, and the (gravitational) lensing profile. We consider the $\Lambda$CDM concordance model of our Universe with $\Omega_m=0.28$, $\Omega_\Lambda=0.72$, the Hubble constant $H_0=100h~$km s$^{-1}$ Mpc$^{-1}$ with $h=0.72$ in our calculations. Dividing the samples into three redshift bins, we find that the LQGs with higher redshift are more biased and correlated than those with lower redshift. The redshift-increasing LQG correlation amplitudes we find is incompatible with that predicted by the standard theory of structure growth. The homogeneity scale $R_H$ of the LQG distribution is also deduced. It is defined as the comoving radius of the sphere inside which the number of LQGs $N(<r)$ is proportional to $r^3$ within $1\%$, or equivalently above which the correlation dimension of the sample $D_2$ is within $1\%$ of $D_2=3$. For the NFW dark matter halo profile, the homogeneity scales of the LQG distribution are $R_H\simeq 224$ $h^{-1}$Mpc for $0.5< z\leq 1$, $R_H\simeq 316$ $h^{-1}$Mpc for $1< z\leq 1.5$, and $R_H\simeq 390$ $h^{-1}$Mpc for $1.5< z\lesssim 2$. These values are above the characteristic sizes of the LQG samples in each bin, implying the validity of the cosmological principle on the LQG scale, i.e. a length range of $200\sim 400~h^{-1}$Mpc and a mass scale of $\sim 10^{14}$M$_\odot$. The possibilities of a top-down structure formation process as was predicted by the hot/warm dark matter (WDM) scenarios and the redshift evolution of bias factor $b$ and correlation amplitude $\xi_{{\rm LQG}}$ of the LQGs as a consequence of the cosmic expansion are both discussed. Theoretical implications as well as the results and observations of other groups are presented.
} 

\section{Introduction}
\label{intro}
Large galaxy redshift surveys have long been carried out to record the angular positions and redshifts of the galaxies. Its primary aim is to measure the matter distribution of our Universe and to determine possible structures on large scales. Since the discovery of CfA2 Great Wall of galaxies \cite{Geller1989}, several large scale structures (LSS) have been reported in the past two decades. The Sloan Great Wall (SGW) \cite{Gott2005} identified in the Sloan Digital Sky Survey (SDSS; \cite{Aihara2011}) is a filamentary structure of galaxies, of which the characteristic size (volume$^{1/3}$) is about $128$ Mpc (unless otherwise stated, all scales presented in this paper are comoving). The U1.11 \cite{Clowes1991} and the U1.28 (or CCLQG) \cite{Clowes2001} are both connected units of quasars, respectively spans about $380$ and $350$ Mpc.

According to the standard structure formation theory, all the structures like galaxies and clusters in our Universe today have their origins in the primordial Gaussian random phase density fluctuations that was generated by the cosmic inflation. The fluctuations of (dark) matter density, once entered the horizon, were amplified by gravity and eventually turned into the rich structures we see today. The validity of this theory has well been supported by a number of observations and $N$-body simulations. One of the inevitable outcomes of this structure formation scenario is the `hierarchical clustering', i.e small systems collapsed first at high redshift in the early Universe (since the wave lengths of the corresponding density fluctuation modes are short and thus entered the horizon at a time earlier than those of long wave lengths), and then grew and merged together to form larger structures like clusters and halos. At the top of this  hierarchical pyramid are the superclusters or the filaments of galaxies/clusters. Combining with the fact that the Universe has a definite age (as $\simeq13.8$ billion years in the $\Lambda$CDM concordance model), one can assume that the structures in our Universe today have a finite size. Yadav et al. \cite{Yadav2010} gave an upper limit of this size as $\sim 260$ $h^{-1}$Mpc (the Hubble constant is given as $H_0=100h~$km s$^{-1}$ Mpc$^{-1}$). They called this the `scale of homogeneity', above which the fractal dimension of the galaxy or matter distribution is equal to the ambient dimension of the space\footnote{In a three-dimensional space, this is `$3$'.}.  

The scale of homogeneity of matter distribution has been studied for more than one decade. The results are quite scattered. Hogg et al. \cite{Hogg2005} once investigated the enormous Luminous Red Galaxy sample \cite{Eisenstein2001} of SDSS \cite{York2000} and presented a homogeneity scale as $R_H\sim 70$ $h^{-1}$Mpc. This result was later supported by Sarkar et al. \cite{Sarkar2009} and Scrimgeour et al. \cite{Scrimgeour2012}, who carried out a multifractal analysis of the distribution of over 200,000 blue galaxies in the WiggleZ survey \cite{Drinkwater2010}. Contrary to these coincidences, Labini et al. \cite{Labini2009} claimed to find a homogeneity scale above $100$ $h^{-1}$Mpc.
Homogeneity on large scales is one of the cornerstones of modern cosmological theory. It assumes that over some large smoothing scale, the distribution of matter is Ôstatistically homogeneousÕ. Small-scale inhomogeneities occur as perturbations of the density field with an isotropic Gaussian distribution. The large-scale homogeneity in the very early Universe (at redshift $z \simeq 1100$) was well supported by the high degree of isotropy of the cosmic microwave background (CMB) radiation power spectrum \cite{Bennett2013} and PLANCK's primary data \cite{Planck2013}. But that does not guarantee the homogeneity of the present Universe. 

Recently, large quasar groups (LQG) that exceed the upper limit of homogeneity scale $\sim 260$ $h^{-1}$Mpc presented by Yadav et al. \cite{Yadav2010} were reported discovered \cite{Clowes2012,Clowes2013}. Quasars, like galaxies and galaxy clusters, are biased tracers of the mass distribution \cite{Kaiser1984}. Unlike galaxies which were mostly formed at $z< 1$, most quasars have redshifts $z > 1$ and thus are suitable candidates to probe the matter distribution of the high-redshift Universe. The clustering of quasars and the underlying dark matter halos have long been studied \cite{Gao2005,Lidz2006,daAngela2008,Steinhardt2010,Rafiee2011}. Unlike the fruitful results about quasars, the clustering of LQGs are rarely discussed in the references. That is because the discovery of a LQG and the determination of their spatial distribution are highly constrained by the insufficient coverage and deepness of the sky by the current sky surveys. No detailed model for the bias of LQG are available so far, although several studies have suggested that their clustering strength is dependent on their bolometric luminosities as well as the mass of the dark matter halos in which they inhabit \cite{Gao2005,Lidz2006,daAngela2008,Steinhardt2010,Rafiee2011}.

In this paper, we extend the investigation of large-scale homogeneity to the scale of LQGs, using the 12 LQGs samples in \cite{Komberg1996}. These groups were identified to have reliable redshifts within the range of $0.5 \lesssim z \lesssim 2$ and thus provide good candidates to explore the mass distribution in the earlier Universe. The fact that quasars being biased tracers of the mass distribution is implemented in that the correlation amplitude of quasars is related to that of the underlying dark matter by a bias factor $b$, i.e. $\xi_{{\rm qua}}=b^2 \xi_{{\rm mass}}$.  
We calculate this factor for the LQGs and study its behavior as a function of $z$ and the halo mass $M_{{\rm DMH}}$ for three different density profiles of the LQG dark matter halos, i.e. the isothermal profile, the NFW profile, and the lensing profile. We consider the $\Lambda$CDM concordance model of our Universe with $\Omega_m=0.28$, $\Omega_\Lambda=0.72$, the Hubble constant $H_0=100h~$km s$^{-1}$ Mpc$^{-1}$ with $h=0.72$ in our calculations. The correlation function $\xi$ of the LQGs and the homogeneity scale $R_H$ are also estimated from the theory. It is based on the `counts-in-spheres' measurement $N(<r)$, i.e. the number of LQGs inside a sphere of radius $r$ centered at each LQG, which is averaged over the whole sample. For an ideal  homogeneous distribution of the LQGs, this number scales as $\propto r^3$ (for $N(<r)={\bar \rho}\frac{4\pi r^3}{3}$, where ${\bar \rho}$ is now a universal constant). The homogeneity scale $R_H$ is defined to be the radius above which this scaling behavior holds with accuracy of $99\%$. This definition of $R_H$ together with the relevant fractal analysis were widely used in the studies of large-scale homogeneity in the galaxy surveys \cite{Hogg2005,Sarkar2009,Labini2009,Yadav2010,Scrimgeour2012} and thus is used in our investigation of LQGs. The results we obtained are compared with those of others and possible explanations are given.

The rest of the paper is organized as follows. In Section 2, we justify the use of the LQG sample in our study. In Section 3, we calculate the correlation function $\xi(r)$ and the bias factor $b$ of the LQG samples from the theory. In Section 3.1, we introduce three different models of bias $b$ for quasars. In Section 3.2, the total halo masses of the LQGs are estimated respectively in three different density profiles of dark matter. In Section 3.3, we give the calculation results of $\xi(r)$ and $b$. The redshift-evolution of them are discussed. The results are plotted. Section 4 are dedicated to study the homogeneity of the LQG distribution. In Section 4.1, we briefly review the definition of fractal dimension and its relation with the homogeneity study. We give in details the criteria of a homogeneous LQG distribution. In Section 4.2, we calculate the homogeneity scale $R_H$ of the LQG distribution in three redshift bins and discuss the $z$-dependence of the results. The correlation dimension as well as the count-in-sphere number of the LQGs as a function of $r$ are plotted in this section.
Conclusions and discussions together with the results and observations of other groups are given in Section 5.

\section{Correlation and bias as a function of redshift}
In this section, we calculate the correlation function $\xi(r)$ and the bias factor $b$ of the LQG sample. We use them to estimate the count-in-sphere number $N(<r)$ and the homogeneous scale $R_H$ for the LQG distribution in the next section.

According to \cite{Peebles1980}, the probability that one can find an object in the volume $dV_1$ with another in $dV_2$ separated by a distance $r$ is $P(r)={\bar n}^2 [1+\xi(r)]dV_1 dV_2$, where ${\bar n}$ is the mean number density.  $\xi(r)$ in the above expression is the two-point correlation function, measuring the clustering degree of the objects. Like that of galaxies and quasars, the correlation function of LQGs is related to that of the underlying dark matter by the bias factor $b$, i.e
\be
\xi_{{\rm LQG}}(r)=b^2 \xi_{{\rm mass}}(r).
\label{LQGcorrel}
\ee
$b$ for the LQGs are regarded to have different values of that for galaxies/clusters or quasars. It is a function of mass and redshift in theory and is usually determined by observations.
The matter correlation function $ \xi_{{\rm mass}}(r)$ is given by the Fourier transform of the power spectrum $P(k)$, i.e.\footnote{Assuming a spherical symmetry, the standard Fourier transform $\xi_{{\rm mass}}(r)\equiv \frac{1}{(2\pi)^3}\int P(k)e^{-i{\mathbf k}\cdot {\mathbf r}}d^3{\mathbf k}$ can be reduced to $\xi_{{\rm mass}}(r)=\frac{1}{2\pi^2}\int P(k)\frac{{\rm sin}kr}{kr}k^2dk$ by integrating over the angle $\theta$ between ${\mathbf k}$ and ${\mathbf r}$.}
\be
 \xi_{{\rm mass}}(r)\equiv\frac{1}{2\pi^2}\int P(k)\frac{{\rm sin}kr}{kr}k^2dk.
 \label{masscorrel}
\ee
 $P(k)\equiv P_0(k) T^2(k)$, where $P_0(k)\equiv (2\pi^2 \delta_H^2k^n/H_0^{n+3})$ is the primordial matter power spectrum generated by the cosmic inflation. $n=1$ is the scalar index number. The factor $\delta_H \simeq 4.6\times 10^{-5}$ is given by the $\Lambda$CDM cosmological model with mass density of $\Omega_{m}=0.28$ and a dark energy density $\Omega_{\Lambda}=0.72$ for nowadays \cite{Dodelson2003}. $T(k)$ is the BBKS transfer function given by \cite{BBKS}\footnote{The BBKS transfer function takes the form as $T(x)=\frac{{\rm ln}(1+0.171x)}{0.171x}\left[1+0.284x+(1.18x)^2+(0.399x)^3+(0.490x)^4\right]^{-0.25}$, where $x\equiv k/k_{{\rm eq}}$ and $k_{{\rm eq}}=0.073~h~{\rm Mpc}^{-1}~\Omega_m h$. $\Omega_m$ is the mass density of the Universe today.}.

\subsection{Three different models of bias}
Despite the ignorance of the bias of the LQG distribution, several models have been proposed for $b$ for quasars. Mo $\&$  White \cite{Mo1996} showed that the bias factor of quasars can be calculated by
\be
b(R,z)=1+\frac{1}{\delta_c}\left[\frac{\delta_c^2}{\sigma_{M}^2(R,z)}-1\right],
\label{bias1}
\ee
where $\delta_c$ is the critical overdensity when the collapse occurs. In a spherical nonlinear collapse model \cite{Padmanabhan2003}, it takes a value of $\delta_c\simeq 1.686$.  $\sigma_{M}(R,z)=\sigma_M(R)/(1+z)$ is called the `rms linear mass fluctuation'\footnote{See Section 9.5 in \cite{Dodelson2003}.} and can be computed by  
\be
\sigma^2_M(R)= \int \frac{d^3k}{(2\pi)^3}P(k)|\tilde{W}(R,k)|^2.
\label{sigma2R}
\ee
$\tilde{W}(R,k)$ is the Fourier transformed top-hat window function\footnote{The window function takes a form as $
\tilde{W}(R,k)=\frac{3}{k^3R^3}\left[-kR{\rm cos}(kR)+{\rm sin}(kR)\right]$.} \cite{Peebles1980}.
$R$ is the so-called `smoothing scale' in the Press-Schechter theory, a formalism developed to describe the full statistical properties of the hierarchical clustering process by Press $\&$ Schechter \cite{Press1974}. It is related to the mass by
\be
R=\left(\frac{M}{4\pi \rho_m}\right)^{1/3}=0.951 ~h^{-1}{\rm Mpc}\left(\frac{M~h}{10^{12}\Omega_{m} ~{\rm M}_\odot}\right)^{1/3},
\label{radius}
\ee
where $\rho_m=\Omega_m \rho_{{\rm cr}}=2.78\times 10^{11}h^2 \Omega_{m}~{\rm M}_\odot {\rm Mpc}^{-3}$. $\rho_{{\rm cr}}$ is the critical density of the Universe. For the LQG samples, we have $M=M_{{\rm DMH}}$, where $M_{{\rm DMH}}$ is the total mass of the dark matter halos of a LQG. The way to estimate $M_{{\rm DMH}}$ is given in the next section.

Besides equation (\ref{bias1}) given by Mo $\&$ White \cite{Mo1996}, other models of the bias factor for quasars have been given. Sheth et al. \cite{Sheth2001} presented an alternative model of $b$, i.e. 
\be
b(\nu,z)&=&1+\frac{1}{\sqrt{a}\delta_c(z)}[\sqrt{a}(a\nu^2)+\sqrt{a}b(a\nu^2)^{1-c}\\
&&-\frac{(a\nu^2)^c}{(a\nu^2)^c+b(1-c)(1-c/2)}],
\label{bias2}
\ee
where $a = 0.707$, $b = 0.5$ and $c = 0.6$. $\nu$ was defined as $\nu\equiv \delta_c(z)/\sigma_{M}^2(R,z)$. The critical density was given as $\delta_c(z)=0.15(12\pi)^{2/3}\Omega_m(z)^{0.0055}$ \cite{NFW1997}. Another model of $b$ was later presented by Mandelbaum et al. \cite{Mandelbaum2005},
\be
b(\nu,z)=1+\frac{q\nu-1}{\delta_c(z)}+\frac{2p}{\delta_c(z)[1+(q\nu)^{p}]},
\label{bias3}
\ee
with $p=0.15$ and $q=0.73$. The definition of $\nu$ is the same as that given in equation (\ref{bias2}). In next section, we will apply these three bias models for quasars to the LQGs to investigate their relations with the underlying distribution of dark matter.

\subsection{Halo masses for LQGs}
To use equations (\ref{bias1}) to (\ref{bias3}) to calculate the bias of LQGs, one has to know how to estimate the total mass of the dark matter halos $M_{{\rm DMH}}$ for a LQG. As the members of a LQG, quasars are usually associated with the active galactic nuclea, of which the emitted energy comes from the accretion process of the black hole in the center. They are often found to inhabit in the dark matter halos with mass $M_{{\rm DMH}}\sim 3\times 10^{12} {\rm M}_\odot$  \cite{Porciani2004,Adelberger2005,Croom2005,Adam2006,Porciani2006}. A redshift-independent relation between the black hole mass $M_{{\rm BH}}$ and that of their hosted dark matter halo is given by \cite{Ferrarese2002} for three different density profiles of dark matter, i.e.
\be
\frac{M_{{\rm BH}}}{10^8 {\rm M}_\odot}\simeq 0.027\left(\frac{M_{{\rm DMH}}}{10^{12} {\rm M}_\odot}\right)^{1.82},~({\rm isothermal ~profile})
\label{isomass}
\ee

\be
\frac{M_{{\rm BH}}}{10^8 {\rm M}_\odot}\simeq 0.1\left(\frac{M_{{\rm DMH}}}{10^{12} {\rm M}_\odot}\right)^{1.65},~({\rm NFW~profile})
\label{NFWmass}
\ee

\be
\frac{M_{{\rm BH}}}{10^8 {\rm M}_\odot}\simeq 0.67\left(\frac{M_{{\rm DMH}}}{10^{12} {\rm M}_\odot}\right)^{1.82}.~({\rm lensing~profile})
\label{lensmass}
\ee

By studying the empirical relation between the central black hole mass in a nearby quasar and its bolometric luminosity\footnote{See the reference \cite{VP2006}.},  Steinhardt et al. \cite{Steinhardt2010} gave a reasonable estimation of the black hole mass as $M_{{\rm BH}}\simeq 10^{8.8} {\rm M}_\odot$ for quasars at redshift $0.5<z\lesssim 2$. Using their results, we invert the equations (\ref{isomass}) to (\ref{lensmass}) to get the mass of the dark matter halo of a LQG, i.e.
\be
M_{{\rm DMH}}=N\cdot \left[\frac{M_{{\rm BH}}}{10^8 {\rm M}_\odot}/0.027\right]^{1/1.82}\simeq 2N\times 10^{13}{\rm M}_\odot ,&&\\
({\rm isothermal ~profile})&& \nonumber
\label{isodmh}
\ee
\be
M_{{\rm DMH}}=N\cdot\left[\frac{M_{{\rm BH}}}{10^8 {\rm M}_\odot}/0.1\right]^{1/1.65}\simeq 1.2N\times 10^{13}{\rm M}_\odot ,&&\\
({\rm NFW ~profile})&& \nonumber
\label{NFWdmh}
\ee
\be
M_{{\rm DMH}}=N\cdot\left[\frac{M_{{\rm BH}}}{10^8 {\rm M}_\odot}/0.67\right]^{1/1.82}\simeq 3.4N\times 10^{12}{\rm M}_\odot ,&&\\
({\rm lensing ~profile})&& \nonumber
\label{lensdmh}
\ee
where $N$ is the number of quasars contained in the group. 
The total halo mass of the LQG samples for the three different dark matter profiles are listed in the column (4), (7) and (10) in Table \ref{table1}.

\subsection{Redshift evolution of LQG bias and correlation}
Given their total halo masses, one can use equations (\ref{bias1}) to (\ref{bias3}) to estimate the bias of LQGs. The NFW density profile of dark matter halos was supported by a number of $N$-body simulation of structure formation \cite{NFW1997} so that it is preferential over the other two profiles. We use the NFW density profile to calculate the mass of the dark matter halo in which a LQG inhabits.
To study the redshift-dependence of the bias factor $b$, the LQG samples are divided into three redshift bins: $0.5< z \leq 1$, $1<z\leq 1.5$, and $1.5<z \lesssim 2$. The mean value of the total halo mass $R$ and the bias factor $b$ are respectively calculated for each bin. The results are listed in Table \ref{table1}.

Figure \ref{fig1} shows the predictions of different bias models for a NFW profile of the LQG halo. The results are presented as a function of $\nu$ to conveniently compare with the results of \cite{Kundic1997} and \cite{Sheth2001} for quasars. It should be noted that the bias model of \cite{Mo1996} predicts a steeper $b(\nu)$ than those predicted from \cite{Sheth2001} and \cite{Mandelbaum2005}. 

Figure \ref{fig2} shows the impacts of different dark halo profiles on the values of $b$. The bias factor $b$ are calculated from the three models \cite{Mo1996}, \cite{Sheth2001}, and \cite{Mandelbaum2005}. We plot the bias factor $b$ as a function of redshift $z$. It shows that the LQGs at higher redshift have a larger mean value of $b$ than those in the low-redshift bin. This implies that the LQGs with higher-redshift are more `biased' than the lower-redshift ones. As the mean total halo masses of these LQGs in the three redshift bins are almost the same (see the column (4), (7) and (10) in Table \ref{table1}), this $z$-increasing bias phenomenon is more like the effect of the cosmic expansion rather than the linear growth of density perturbation. Similar result was also obtained from a $N$-body simulation for halos less massive than $\sim 10^{13}$ M$_\odot$ by \cite{Gao2005}.

The LQG correlation functions (i.e. the equation (\ref{LQGcorrel})) for different dark halo density profiles are presented in Figure \ref{fig3}. The isothermal density profile, which has larger $M_{{\rm DMH}}$ values, has higher correlation amplitudes than the other two profiles. It means that the more massive the LQGs are, the more correlated they are. They all show an apparent $z$-increasing correlation amplitudes as that reported by Kundic \cite{Kundic1997} for quasars. Similar results have been obtained by Franca et al. \cite{Franca1998} and Gao et al. \cite{Gao2005}. These facts are inconsistent with the prediction from the linear theory of structure growth that the overdensity $\delta$ scales as\footnote{See equation (8.37) in Section 8.2 of \cite{Padmanabhan2003}.} $\propto (1+z)^{-1}$. Possible explanations and physical implications of this are discussed in the last section.

\section{Homogeneity of the LQG distribution}
Given the correlation function (\ref{LQGcorrel}) of the LQGs, we can now estimate the homogeneity scale of the LQG samples.

\subsection{Correlation dimension $D_2(r)$ for LQGs}
Before we study the homogeneity of the LQG distribution, we had better properly define what we mean by the so-called `homogeneity scale'. A number of methods have been developed to study the homogeneity of the galaxy distribution, prominent among them being the fractal analysis \cite{Yadav2005}. Fractal refers to a geometrical object that every small part of it appears as a reduce of the whole. In fractal analysis, the concept `fractal dimension' was invoked to describe the homogeneity of the distribution of a point set. It can be related to the $n$-point correlation function. One of the definitions of fractal dimension which are used most often is the `correlation dimension' $D_2(r)$. It is used to quantify the scaling behavior of the two-point correlation function of a sample.

In this paper, we use the correlation dimension to study the homogeneity of the LQGs distribution. We adopt the working definition presented by Scrimgeour et al. \cite{Scrimgeour2012} for a finite point set. It is based on the `counts-in-spheres' measurement. Given a set of points in space, the measurement is to find the average number $N(<r)$ of neighbouring points inside a sphere of radius $r$ centered at each point. If the distribution of points is homogeneous, this quantity should scale as $N(<r)=4\pi r^3 {\bar n}/3\propto r^3$, where ${\bar n}$ is the mean number density of points in that region. To formulate this, suppose we take a LQG in the sample and count the number of its neighbors within a distance $r$. The number scales as
\be
N(<r)\propto r^D,
\ee
The factor $D$ is the correlation dimension.
As was mentioned in the introduction, for a homogeneous LQG distribution, we will have $D= 3$, which equals the ambient dimension (the dimension of the neighboring space). If $D < 3$, the LQG samples are considered to have an inhomogeneous distribution pattern. If $D > 3$, the distribution is called `super-homogeneous' \cite{Gabrielli2002}.
In references, $N(<r)$ is usually corrected for incompleteness by dividing by the number expected for a homogeneous distribution. This gives
\be
{\mathcal N}(<r)\propto r^{D-3}.
\ee
Under this correction, the criteria for homogeneity mentioned above still holds, but now taking a form as ${\mathcal N}(<r)\propto r^{3-3}=1$. The correlation dimension $D$ is actually a function of the sphere radius $r$ and is often written as $D_2(r)$ in three-dimensional space. It can be deduced from the count-in-sphere number $N(<r)$ as
\be
D_2(r)\equiv \frac{{\rm d}~ln N(<r)}{{\rm d}~ln r}=\frac{{\rm d}~ln {\mathcal N}(<r)}{{\rm d}~ln r}+3.
\label{D2}
\ee
The correlation dimension $D_2(r)$ measures the scaling properties of ${\mathcal N}(<r)$ without being dependent on its amplitude (the later of which is affected by the mean number density of LQGs in the regions.) Thus, it is an objective and reliable measure of homogeneity in the statistical analysis.

To calculate the correlation dimension $D_2(r)$ of the LQG sample from the theory,  one has to know the relation between $N(<r)$ and the two-point correlation function of the LQGs, i.e. $\xi_{{\rm LQG}}$.  This is given by \cite{Peebles1980} 
\be
N(<r)={\bar n}\int_0^r [1+\xi_{{\rm LQG}}(r^\prime,z)]4\pi r^{\prime 2}dr^\prime.
\label{Nfromxi}
\ee
$\xi_{{\rm LQG}}(r,z)$ can be estimated from equation (\ref{LQGcorrel}), given the correlation function (\ref{LQGcorrel}) and bias $b$ of the LQGs. These have been calculated in the last section. To correct for the incompleteness, we divide the counts-in-spheres prediction $N(<r)$ in equation (\ref{Nfromxi}) by the number expected for a random distribution, i.e. ${\bar n}\frac{4}{3}\pi r^3$, to get
\be
{\mathcal N}(<r)=\frac{3}{4\pi r^3} \int_0^r [1+\xi_{{\rm LQG}}(r^\prime)]4\pi r^{\prime 2}dr^\prime.
\label{N2}
\ee
The scaled count-in-sphere number ${\mathcal N}(<r)$ and the correlation function $D_2(r)$ for the LQG samples are plotted in Figure \ref{fig4} and Figure \ref{fig5}.

\subsection{Scale of homogeneity $R_H$}
There are several ways to define the homogeneity scale $R_H$ with respect to the correlation dimension $D_2(r)$. \cite{Yadav2010} set the line of homogeneity scale to be the scale above which the deviation of the fractal dimension $D_2(r)$ from the ambient dimension becomes smaller than the statistical dispersion of $D_2(r)$. For the small number of the LQG samples, the statistical dispersion of $D_2(r)$ by this definition might be large. Thus, in our analysis, we use a more robust definition of $R_H$ which is not affected by the sample size. This definition of $R_H$ was presented by \cite{Scrimgeour2012}. In their paper, the scale of homogeneity was defined as the scale above which the correlation dimension $D_2(r)$ of the sample is within $1\%$ of $D_2=3$, i.e. $D_2(r=R_H)=2.97$. An equivalent definition of $R_H$ can also be given as the comoving radius of the sphere inside which the number of LQGs $N(<r)$ is proportional to $r^3$ within $1\%$, i.e. ${\mathcal N}(r=R_H)=1.01$. Under these definitions, we calculate the homogeneity scale of the distribution of the LQGs sample. A factor of 1.55 is multiplied by $P(k)$ in equation (\ref{masscorrel}) to restore the redshift distortion of the observed power spectrum. 

We consider the bias model given by Mo $\&$ White \cite{Mo1996} as in equation (\ref{bias1}) in our calculations. Since the NFW density profile of dark matter halos was supported by a number of $N$-body simulation of structure formation \cite{NFW1997}, it is taken preferential over the other two profiles. The conclusions and discussions in this paper refer mainly to the values of $R_H$ that calculated considering a NFW profile of the LQG halos. 

Theoretical predictions of ${\mathcal N}(<r)$ and $D_2(r)$ as well as the corresponding value of $R_H$ for the LQG samples are shown in Figure \ref{fig4} and Figure \ref{fig5}.
The solid lines are calculated from equations (\ref{N2}) and (\ref{D2}), considering the mean bias value of the LQGs in each bin (see the row titled `Mean' in the column (6), (9) and (12) in Table \ref{table1}. The specific values of $R_H$ for the three dark halo profiles are listed in Table \ref{table4}.

For the NFW profile and the bias model of \cite{Mo1996} in equation (\ref{bias1}), the results are $R_H\simeq 224$ $h^{-1}$Mpc for $0.5< z\leq 1$, $R_H\simeq 316$ $h^{-1}$Mpc for $1< z\leq 1.5$, and $R_H\simeq 390$ $h^{-1}$Mpc for $1.5< z\lesssim 2$. The LQG samples all have characteristic sizes that are under the homogeneity scale $R_H$ in that bin. This fact manifests that the LQG samples identified by \cite{Komberg1996} are consistent with the cosmological principle, the assumption of homogeneous matter distribution after smoothing over some large scale, i.e. $R_H$.
It should also be noted that the homogeneity scale $R_H$ of the LQG distribution shows an increasing dependence on the redshift $z$ like the bias $b$ and correlation amplitude $\xi$. Discussions about this is given in the next section. The characteristic sizes of the LQG samples as well as other large-scale structures are shown in Figure \ref{lqgs} to compare with the theory-predicted $R_H$, with the data listed in Table \ref{table2} and Table \ref{table3}.

\section{Conclusions and Discussion}
In this paper, we investigated the homogeneity of the matter distribution in the high-redshift universe. We based our study on the 12 high-quality LQG samples in \cite{Komberg1996}. We calculated the correlation function $\xi(r)$ and the bias factor $b$ of the LQG samples and studied their dependence on the halo mass $M_{{\rm DMH}}$ and the redshift $z$. The result were presented in three different density profiles of $M_{{\rm DMH}}$. We considered the $\Lambda$CDM concordance cosmological model with $\Omega_m=0.28$, $\Omega_\Lambda=0.72$, the Hubble constant $H_0=100h~$km s$^{-1}$ Mpc$^{-1}$ with $h=0.72$ in our calculations. The LQGs were divided into three redshift bins as $0.5<z\leq1$, $1<z\leq 1.5$ and $1.5<z\lesssim 2$. A $z$-increasing behavior of the bias factor $b$ of the LQG samples has been found. 
The correlation function amplitudes of the LQG samples in the high-redshift bins were also found to have larger values than those in the low-redshift bins. Similar $z$-increasing correlation behaviors have been found in quasars and halos by other groups. To name a few, Kundic \cite{Kundic1997} reported that there was a `clustering excess' between the quasar samples with $z>2$ and those with $z<2$, which means that the amplitude of the quasar correlation function of the high-redshift samples was significantly higher than that of the low-redshift sample. This result was later supported by \cite{Franca1998}, who claimed to find a weak $z$-increasing quasar-quasar correlation from $z<1.4$ to $z>1.4$. Similar results were also reported in the investigations of the clustering of dark matter halos less massive than $\sim 10^{13}~h^{-1}{\rm M}_\odot$ \cite{Gao2005}.

All these facts implied that the early Universe was more biased than the nearby or present one. This scenario is incompatible with the current structure formation theory which predicts a mass overdensity perturbation growth rate scaling as $\delta \propto (1+z)^{-1}$ \cite{Padmanabhan2003}.
Moreover, it should also be noted that the LQG samples in these three different redshift bins have almost the same characteristic mean size as well as the dark halo masses, although their bias $b$ and the correlation function amplitude $\xi$ are showing a $z$-increasing trend. This implies that the $z$-evolution of $b$ and $\xi$ of LQGs are more like a result of the cosmic expansion than that of the nonlinear growth of the density perturbation. At very early times, the Universe was matter-dominated in which the density perturbation grew with time. As the dark energy set in and predominated at late times, the Universe began to expand at an accelerated rate. This phase transition has a non-dismissible impact on the the time evolution of the structures in the late Universe. Incidentally, the accelerated cosmic expansion takes place at $z \gtrsim 0.4$ (this can been seen from the fact that the dark energy density $\Omega_{\rm \Lambda}$ was strongly influenced by the supernovae samples with redshift $0.4 \lesssim z \lesssim 0.8$ \cite{Riess1998,Perlmutter1999}), which is coincident with the redshift range $0.5<z \lesssim 2$ we studied in this paper. Further investigations would be necessary to find out whether the nonlinear growth of the density perturbation could be overtaken by the expansion of the Universe. That would be too much to be included in this paper.

Two features of our investigation are worth mentioned. First, we studied the homogeneity of matter clustering on a mass scale over $\sim 10^{13}$M$_\odot$, while the previous investigations were mainly on the mass range $< 10^{13}$M$_\odot$ and for galaxies or quasars. Second, we used the LQGs to study the large-scale homogeneity of the high-redshift Universe, i.e. $0.5<z\lesssim2$, compared with those results obtained from galaxy or quasars investigations within the redshift range $z<1$.

Given the correlation function of the LQG samples, the scale of homogeneity defined by fractal dimension was also estimated for the three different dark halo density profiles from the theory. The results were illustrated in Figure \ref{N2} and \ref{D2}. The homogeneity scale $R_H$ obtained for the three different redshift bins show a $z$-increasing behavior as that of the correlation function $\xi$ and the bias $b$. The $z$-increasing $R_H$ values indicates that the early Universe has a larger region of inhomogeneous matter clustering such as the galaxies or clusters formation than that in a nearby Universe. This scenario is compatible with that implied by a $z$-increasing $b$ and $\xi$. 
The results of $R_H$ together with the characteristic size of the LQG samples were shown in Figure \ref{lqgs}. Corresponding data were listed in Table \ref{table2}. 
The results we obtained for the redshift bin $0.5<z\leq 1$ for three density dark halo profiles are consistent with the upper limit $\sim 260~h^{-1}$Mpc that placed by \cite{Yadav2010} as well as the characteristic size of the LQG samples in that bin. For the bins $1<z\leq 1.5$ and $1.5<z\lesssim 2$, the values we obtained are higher than that given by \cite{Yadav2010} but are still compatible with the characteristic sizes of the LQG samples. The fact that the scale of the characteristic size of the 12 LQG samples are all under the homogeneity scale $R_H$ in that bin for three density profiles implies that the cosmological principle still holds on the LQG scale, i.e. the redshift range of $0.5<z\lesssim 2$ and a mass scale of $\sim 10^{14}$ M$_\odot$. 

As a matter of fact, the above conclusions also hold for other LQGs and large-scale structures like SGW and CfA2 GW that are listed in Table \ref{table3}. The only exception is the Huge-LQG.
The Huge-LQG, discovered by \cite{Clowes2013}, is a LQG that consists of 73 quasars and has a characteristic size of $\sim 360~h^{-1}$Mpc (equivalently $\sim 500$ Mpc for $h=0.72$). It is slightly larger (if not considered marginally compatible with) than the upper size of structures that are possibly allowed in the current structure formation theory. 
Possible implication is that the discovery of `abnormally' large LQGs such as the Huge-LQG favors an isothermal density profile of dark matter halos. More high-quality data of LQG are required for a further study about this issue. Since the identification of a LQG or large structure is not an easy task, we leave it for future investigations.

Although the diverse data we used were insufficient for a concrete conclusion, there were signs 
that our Universe was once more clustered than it is nowadays. Thomas et al. \cite{Thomas2011} studied the lowest multipoles in the angular power spectra $C_{\ell}$ in the redshift range $0.45\leq z \leq 0.65$ and reported observing a large excess of power in the clustering of luminous red galaxies in the MegaZ DR7 from the SDSS galaxy sample. Similar result was earlier obtained by \cite{Busswell2004}, who reinvestigated data of the 2dF Galaxy Redshift Survey (2dFGRS) and claimed that the existence of any void in the galaxy distribution would result in excess power in the two-point correlation function of galaxies on large scales. All these results, as concluded by \cite{Labini2011} in their paper, were inconsistent with the standard excursion set theory for structure growth in a concordant $\Lambda$CDM Universe. 
A number of possible explanations were proposed, such as modified gravity, exotic dark energy, redshift distortions, large-scale inhomogeneity as was suggested in this paper, or non-Gaussianity, etc. An interesting conception is that the Universe is multilayer-structured, like that of a polystyrene plastic foam lunch box. This can in principle be implemented by an oscillating primordial power spectrum $P(k)$, which provided the seeds of inhomogeneities for the structure formation in the early Universe. Signs of resonance in the CMB power spectrum was recently reported by \cite{Meerburg2014} and \cite{Miranda2014}. 
Whether these could be the smoking gun of new physics over the largest scales probed by the survey nowadays, i.e. during the dark age of the Universe, are still subject to further investigations. 
A top-down structure formation process was also one of the theoretical predictions of the hot/warm dark matter (WDM) hypothesis. The most common WDM candidates are sterile neutrinos\footnote{Sterile neutrinos, also called inert neutrinos, are hypothetical particles that do not interact with other particles via any of the fundamental interactions in the Standard Model of particle physics except gravity. This term usually also refers to neutrinos with right handed chirality \cite{Drewes2013}.} and gravitinos\footnote{Gravitino is the supersymmetric partner of the graviton, which is predicted by the supergravity theories, the theories which combines general relativity and supersymmetry.}. In the WDM scenario, free-streaming of the WDM particles in the very early Universe (i.e. the Universe with temperature $T\simeq 1$ MeV) would ``wash away'' the mass overdensity on small and medium scales (i.e. the scale of that of galaxy and galaxy clusters), leaving only the overdensity on large scales (i.e. above the scale of that of superclusters) to grow in the late times. This would result in a top-down structure formation process in the late-time Universe, in which structures on large scales were first formed and grew and then splitted into several substrutures on small scales. This scenario is incidentally consistent with the discovery of super-structures like Huge-LQG on large scales, the size of which was found to exceed the upper limit predicted by the current linear structure formation theory. 
WDM models \cite{Dudas2014,Ishida2014} have also been found to be able to account for the recently discovered emission line at about $\simeq 3.5 $ keV in the X-ray spectrum of galaxy clusters \cite{Boyarsky2014,Bulbul2014}.
 
Objects like quasars and LQGs and events such as gamma ray bursts usually have higher redshifts than galaxies and thus provide new tools to study the matter clustering and any possible power excess in the early Universe. Provisional conclusion, however, is that the next generation of sky surveys offer excellent prospects for constraining the current model and theory of the structure formation as well as the cosmological principle. Relevant research and observations are currently undertaking.

\section*{Acknowledgments}
We are grateful to Jun-Hui Fan from the Guangzhou University, Shuang-Nan Zhang, Chang Zhe, and Hai-Nan Lin from the Institute of High Energy Physics for useful discussions. We thank Hao Liu from the theory division of the Institute of High Energy Physics for her help during the author's visit in the institute.

%
%

\newpage

\begin{figure} 
\centering
\subfigure[~\textsf{$b(\nu)$ for isothermal profile}] { \label{fig1a}
\scalebox{0.33}[0.33]{\includegraphics{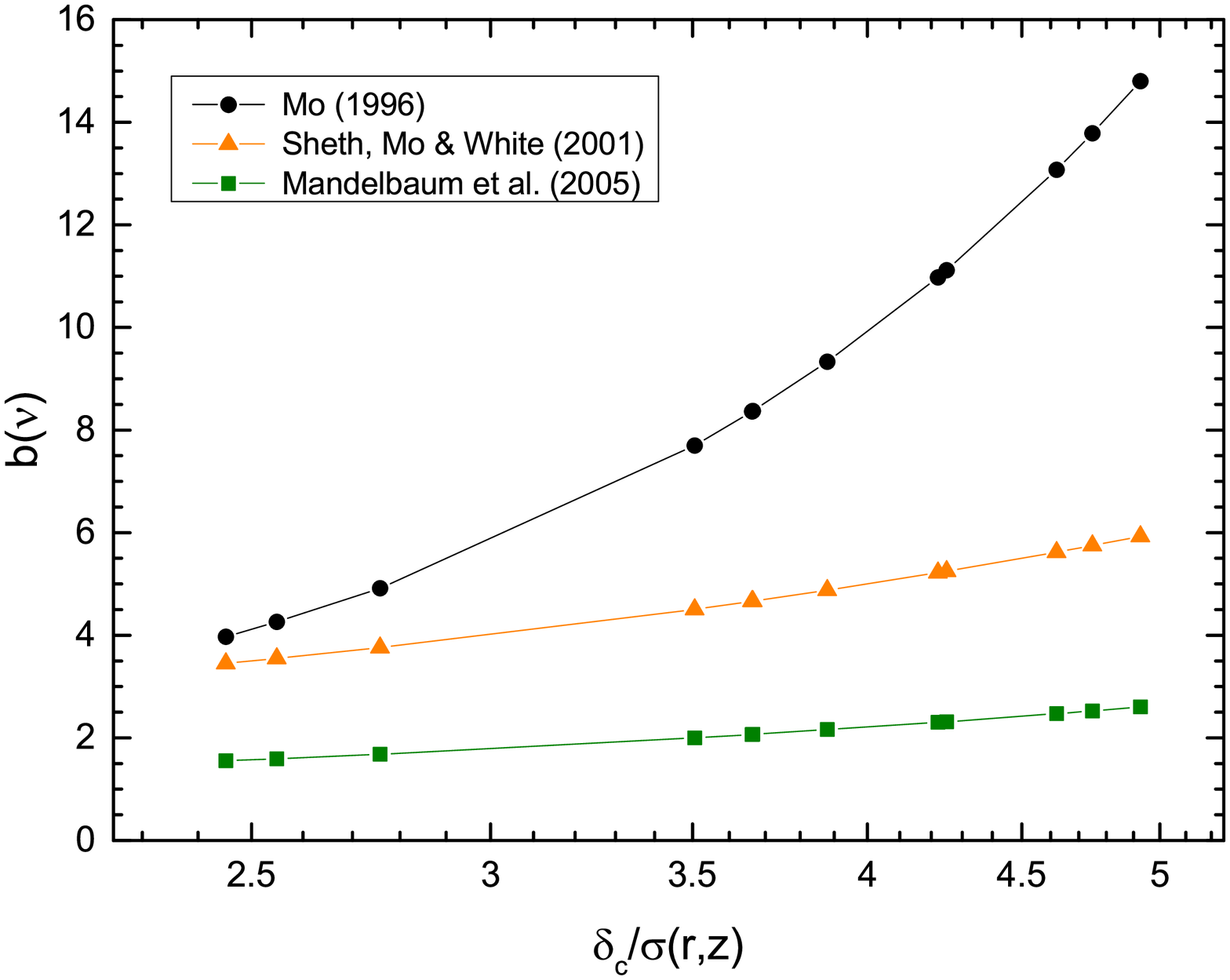}}
}
\subfigure[~\textsf{$b(\nu)$ for NFW profile}] { \label{fig1b}
\scalebox{0.33}[0.33]{\includegraphics{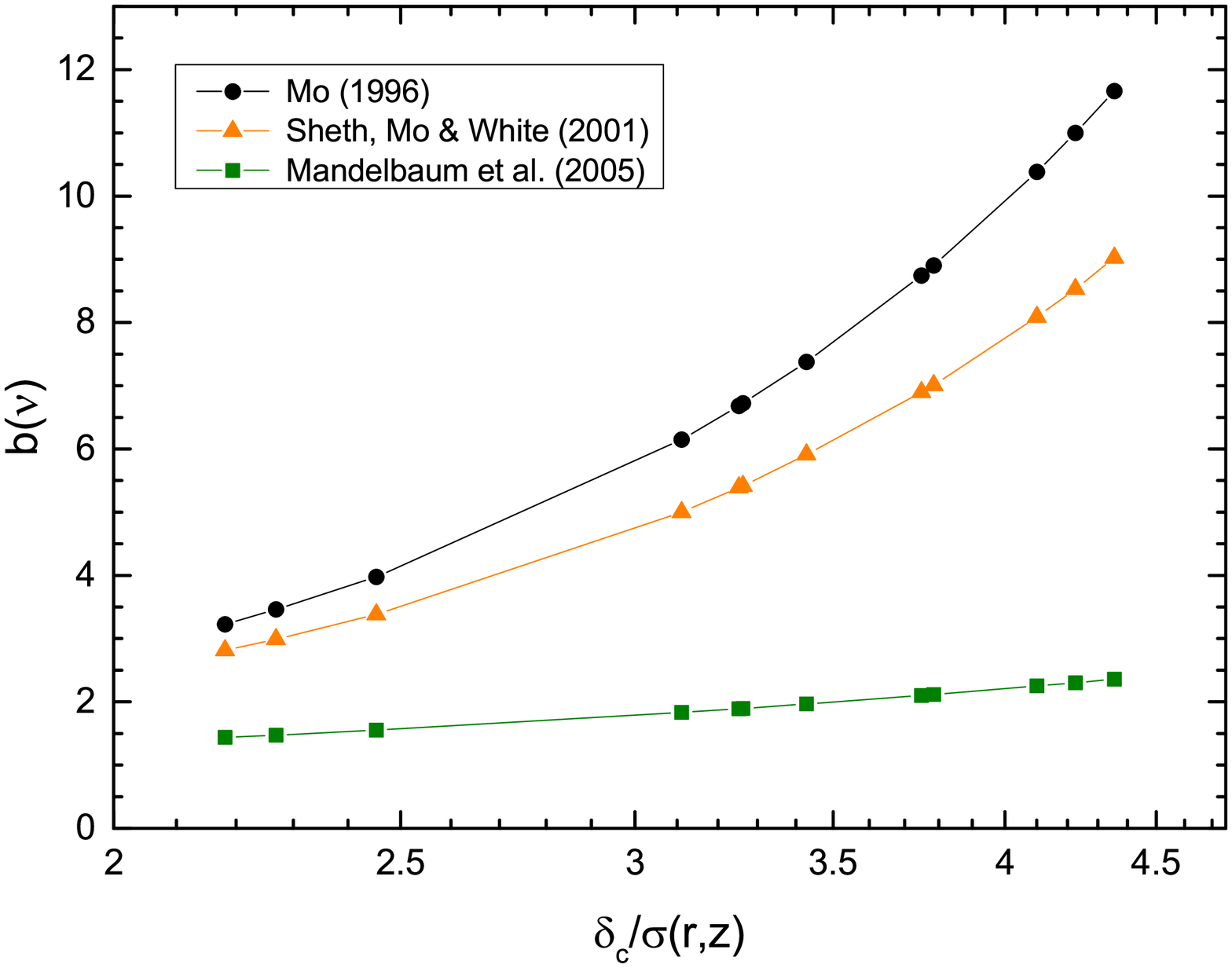}}
}
\subfigure[~\textsf{$b(\nu)$ for lensing profile}] { \label{fig1c}
\scalebox{0.33}[0.33]{\includegraphics{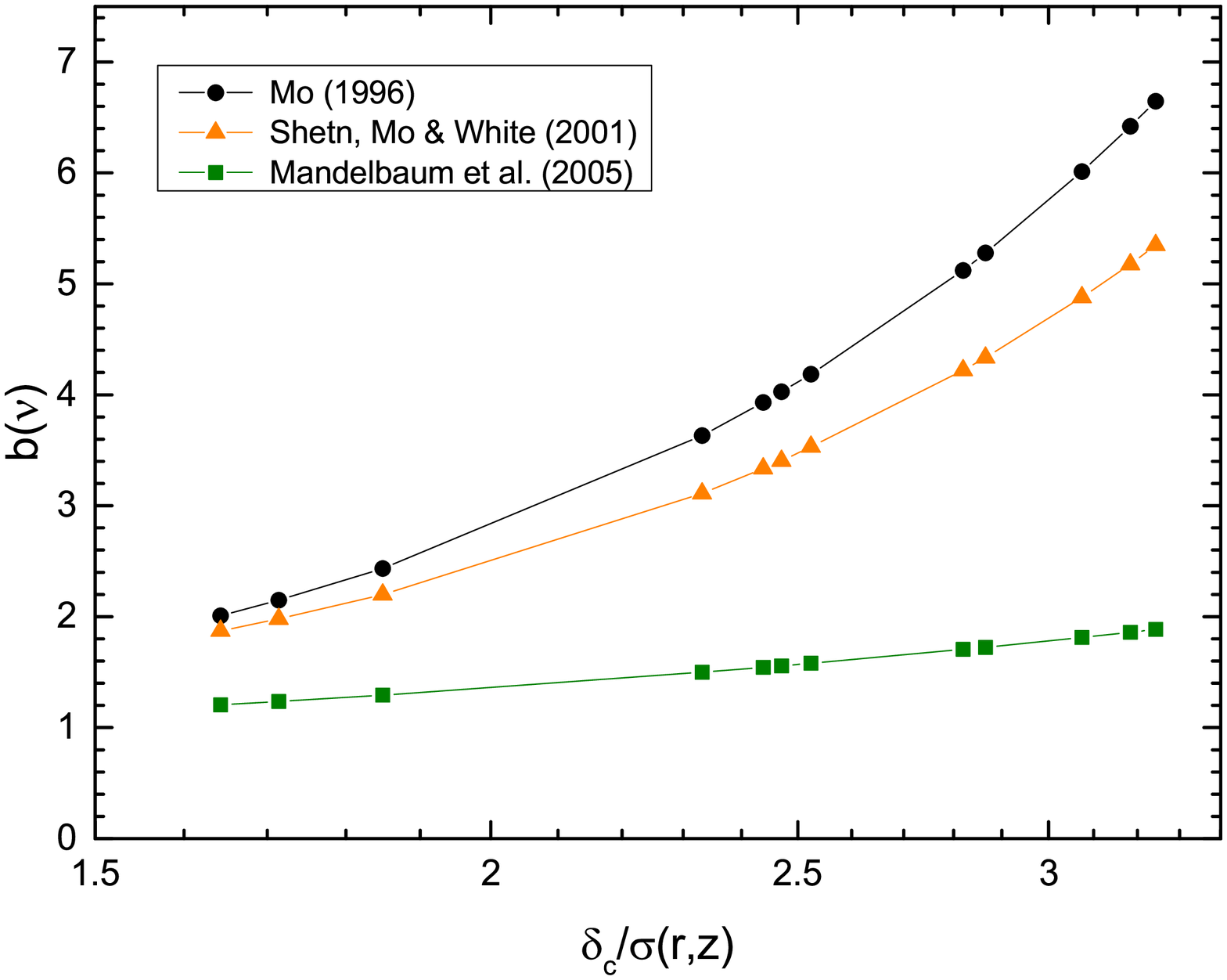}}
}
 \caption{Different models of bias $b(\nu)$ for the three halo profiles. The $x$-axis is given in $\nu=\delta_c/\sigma(R,z)$. Circles with lines indicate the results calculated by equation (\ref{bias1}) from Mo $\&$ White [28]. The cyan and orange lines with symbols are respectively the predictions from Sheth et al. [31] and Mandelbaum et al. [33], i.e. the equation (\ref{bias2}) and (\ref{bias3}). (a), (b) and (c) are in sequence the predictions for the isothermal profile, the NFW profile, and the lensing profile.}
 \label{penG}
\end{figure}

\begin{figure*}
\centering
\subfigure[~\textsf{$b(z)$ of Mo $\&$ White (1996)}] { \label{fig2a}
\scalebox{0.33}[0.33]{\includegraphics{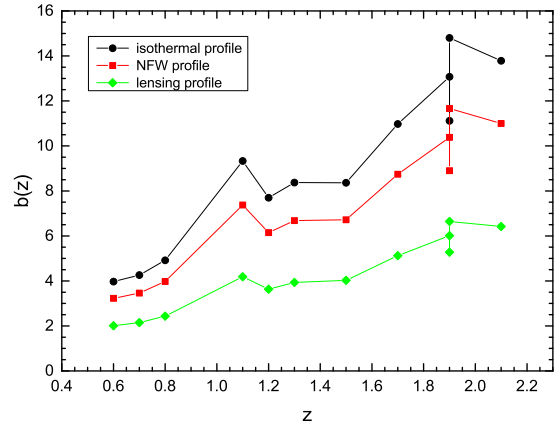}}
}
\subfigure[~\textsf{$b(z)$ of Sheth et al. (2001)}] { \label{fig2b}
\scalebox{0.33}[0.33]{\includegraphics{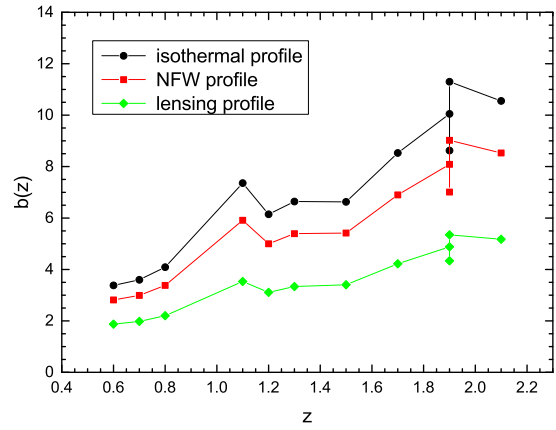}}
}
\subfigure[~\textsf{$b(z)$ of Mandelbaum et al. (2005)}] { \label{fig2c}
\scalebox{0.33}[0.33]{\includegraphics{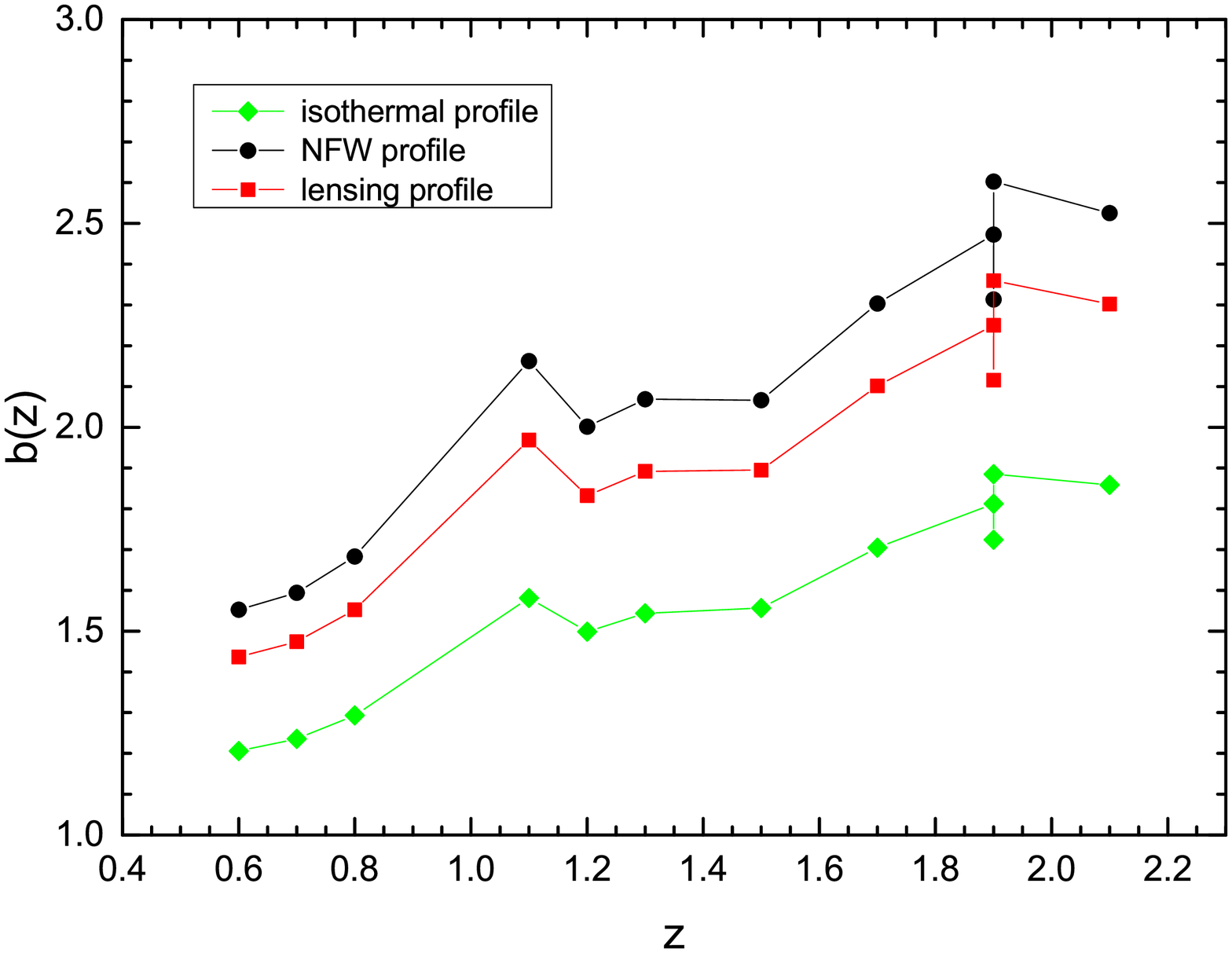}}
}
 \caption{\small{Bias factor $b(z)$ for different density profiles of dark halos, shown in three different models of $b$. Circles, rectangles and diamonds with solid lines in sequence represent the analytical predictions from for the isothermal, the NFW, and the lensing profile of the dark matter halos of the LQGs.  (a), (b) and (c) are in sequence the predictions from  Mo $\&$ White [28] (equation (\ref{bias1})), Sheth et al. [31] (equation (\ref{bias2})), and Mandelbaum et al. [33] (equation (\ref{bias3})).
}}
\label{fig2}
\end{figure*}

\begin{figure*}
\centering
\subfigure[~\textsf{$\xi_{{\rm LQG}}(r)$ for isothermal profile}] { \label{fig3a}
\scalebox{1}[1]{\includegraphics{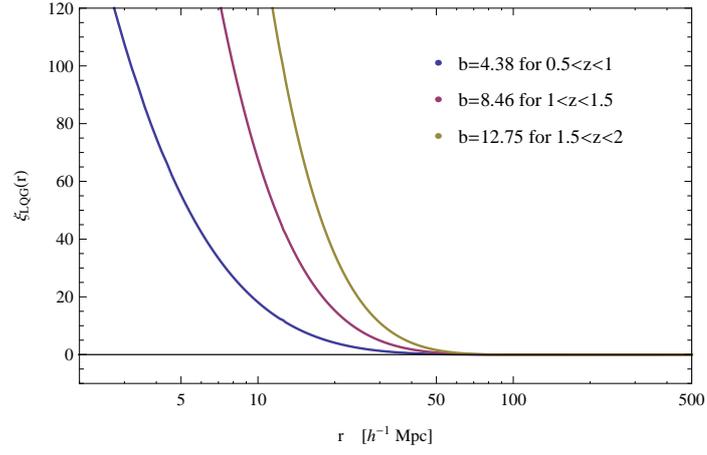}}
}
\subfigure[~\textsf{$\xi_{{\rm LQG}}(r)$ for NFW profile}] { \label{fig3b}
\scalebox{1}[1]{\includegraphics{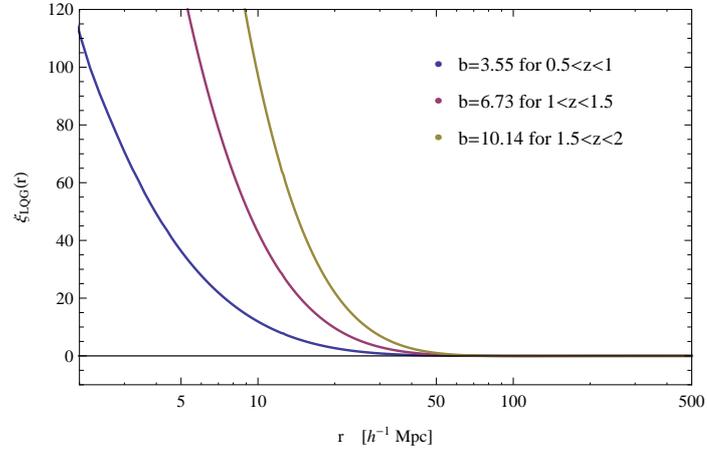}}
}
\subfigure[~\textsf{$\xi_{{\rm LQG}}(r)$ for lensing profile}] { \label{fig3c}
\scalebox{1}[1]{\includegraphics{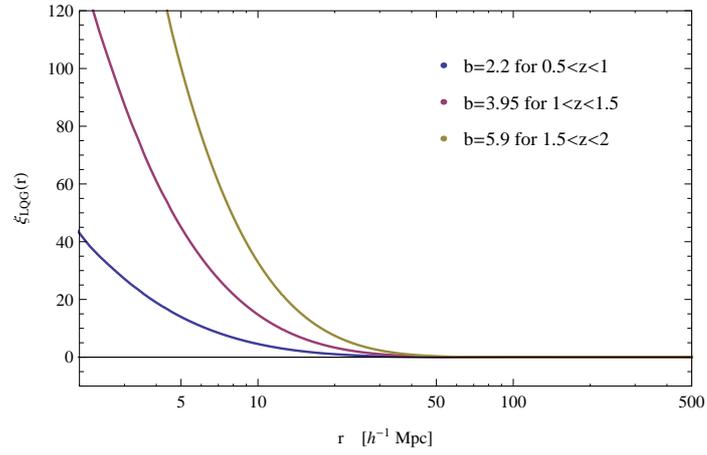}}
}
 \caption{\small{LQG correlation function $\xi_{{\rm LQG}}(r)$ for different dark halo density profiles. (a), (b) and (c) are in sequence the results for the isothermal profile, the NFW profile, and the lensing profile. Solid lines from the bottom to the top are in sequence the calculation results from the equation (\ref{LQGcorrel}) for the mean bias value in the three redshift bin, i.e. $0.5< z\leq 1$, $1< z\leq 1.5$ and $1.5< z\leq 2$.}}
\label{fig3}
\end{figure*}

\begin{figure*}
\centering
\subfigure[~\textsf{${\mathcal N}(<r)$ for isothermal profile}] { \label{fig4a}
\scalebox{1}[1]{\includegraphics{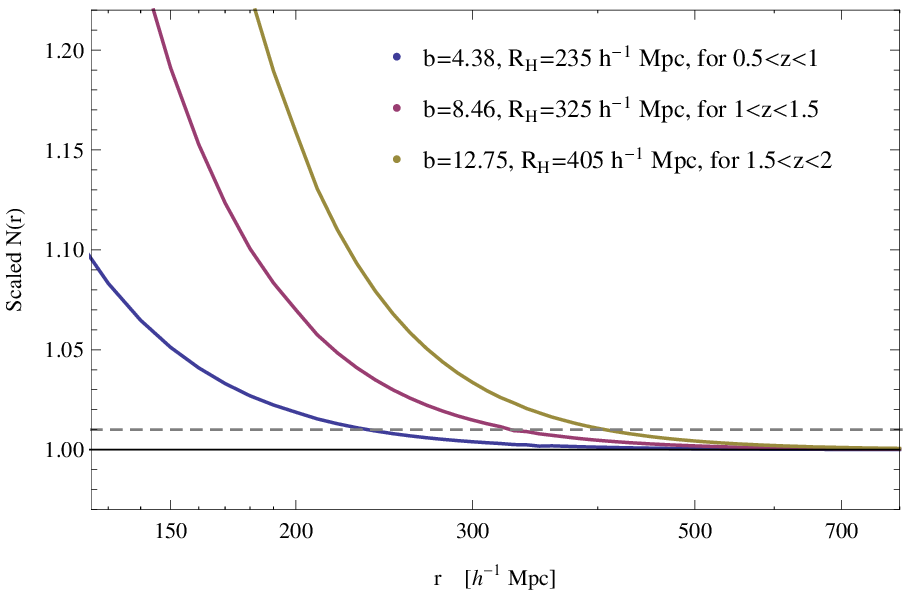}}
}
\subfigure[~\textsf{${\mathcal N}(<r)$ for NFW profile}] { \label{fig4b}
\scalebox{1}[1]{\includegraphics{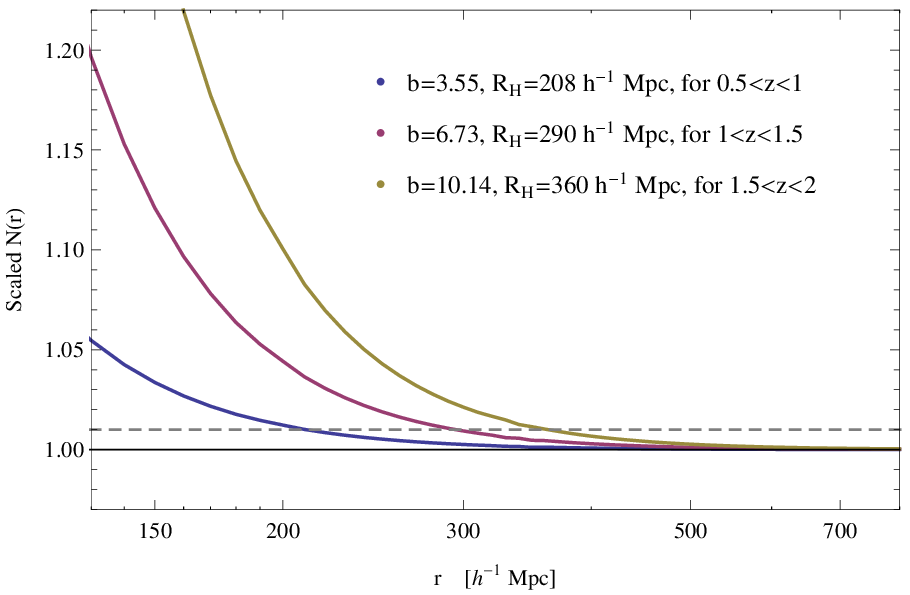}}
}
\subfigure[~\textsf{${\mathcal N}(<r)$ for lensing profile}] { \label{fig4c}
\scalebox{1}[1]{\includegraphics{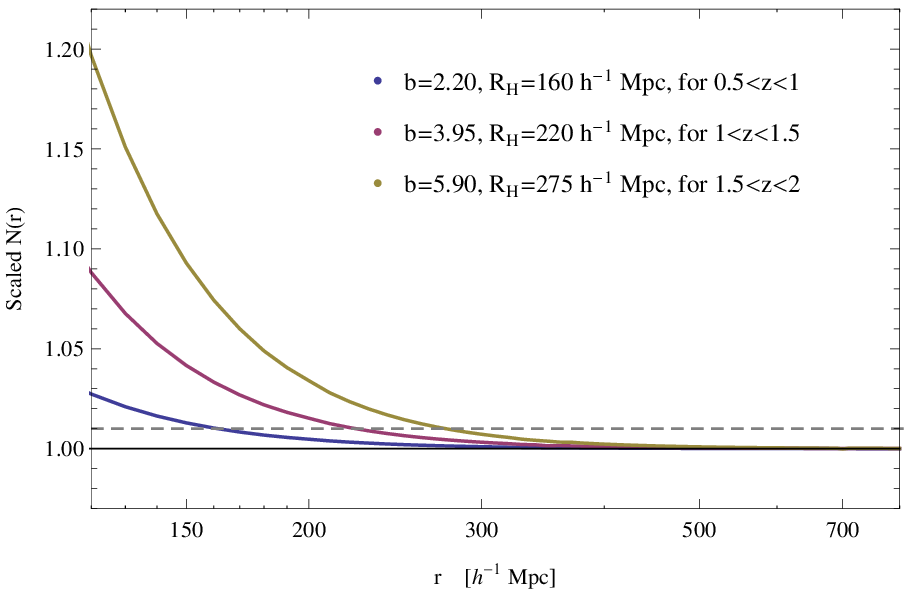}}
}
 \caption{\small{Scaled count-in-sphere ${\mathcal N}(<r)$ of the LQGs for different dark halo density profiles. Solid lines from the left to the right are in sequence the predictions of equation (\ref{N2}) for the three redshift bins in each profile, i.e. $0.5<z\leq 1$, $1<z\leq1.5$ and $1.5<z\leq 2.1$. The results are calculated using the mean bias value of the LQGs in that bin (see the row titled `Mean' in the column (6), (9) and (12) in Table \ref{table1}. The dashed indicates the critical value defined for the transition from a homogeneous to an inhomogeneous distribution of the LQGs, e.g. 1$\%$ from the homogeneity, ${\mathcal N}(<r)=1.01$. The corresponding homogeneity scale $R_H$ are shown in legends.}}
\label{fig4}
\end{figure*}

\begin{figure*}
\centering
\subfigure[~\textsf{$D_2(r)$ for isothermal profile}] { \label{fig5a}
\scalebox{1}[1]{\includegraphics{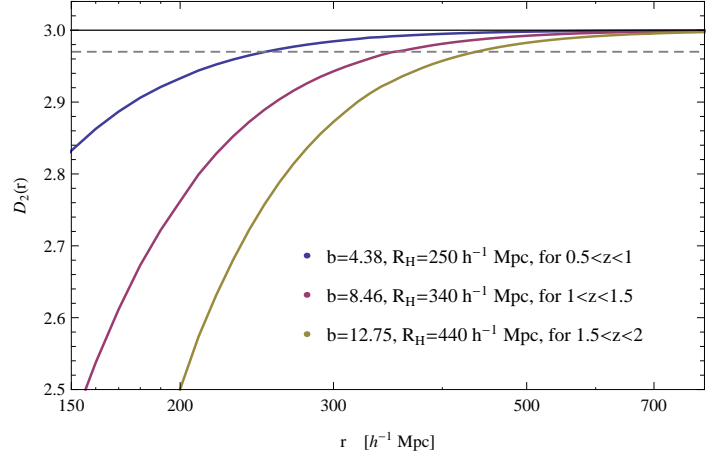}}
}
\subfigure[~\textsf{$D_2(r)$ for NFW profile}] { \label{fig5b}
\scalebox{1}[1]{\includegraphics{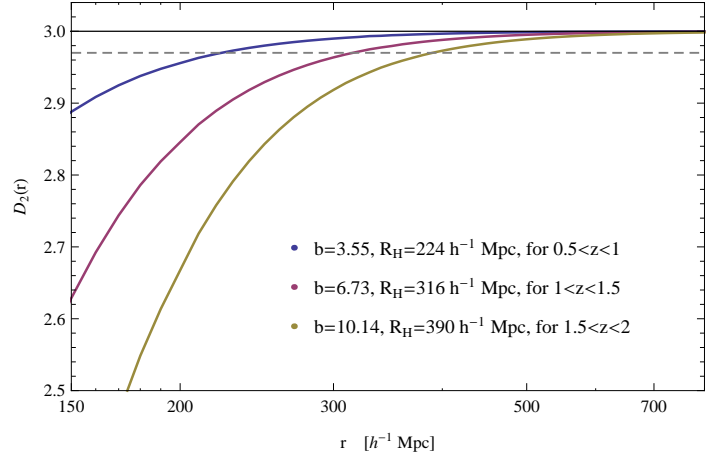}}
}
\subfigure[~\textsf{$D_2(r)$ for lensing profile}] { \label{fig5c}
\scalebox{1}[1]{\includegraphics{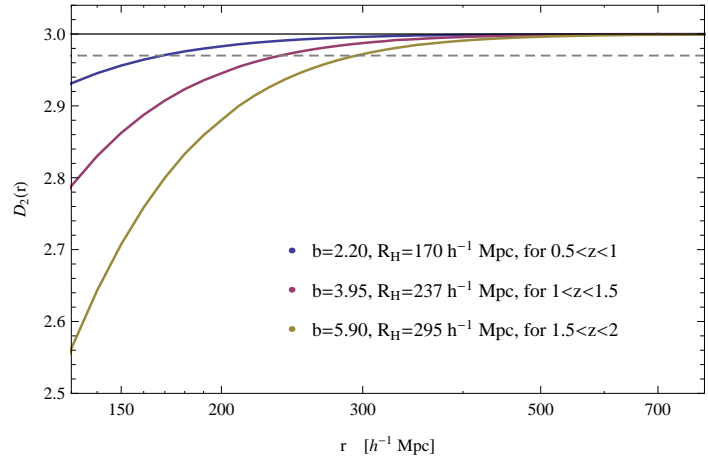}}
}
 \caption{\small{Correlation dimension $D_2(r)$ of the LQGs for different dark halo density profiles. Solid lines from the left to the right are in sequence the predictions of equation (\ref{D2}) for the three redshift bins in each profile, i.e. $0.5<z\leq 1$, $1<z\leq1.5$ and $1.5<z\leq 2.1$. The results are calculated using the mean bias value of the LQGs in that bin (see the row titled `Mean' in the column (6), (9) and (12) in Table \ref{table1}. The dashed indicates the critical value defined for the transition from a homogeneous to an inhomogeneous distribution of the LQGs, e.g. 1$\%$ from the homogeneity, $D_2(r)=2.97$. The corresponding homogeneity scale $R_H$ are shown in blue. The results are consistent with that given in Figure \ref{fig4}.}}
\label{fig5}
\end{figure*}

\begin{figure*}
 \includegraphics[scale=0.6]{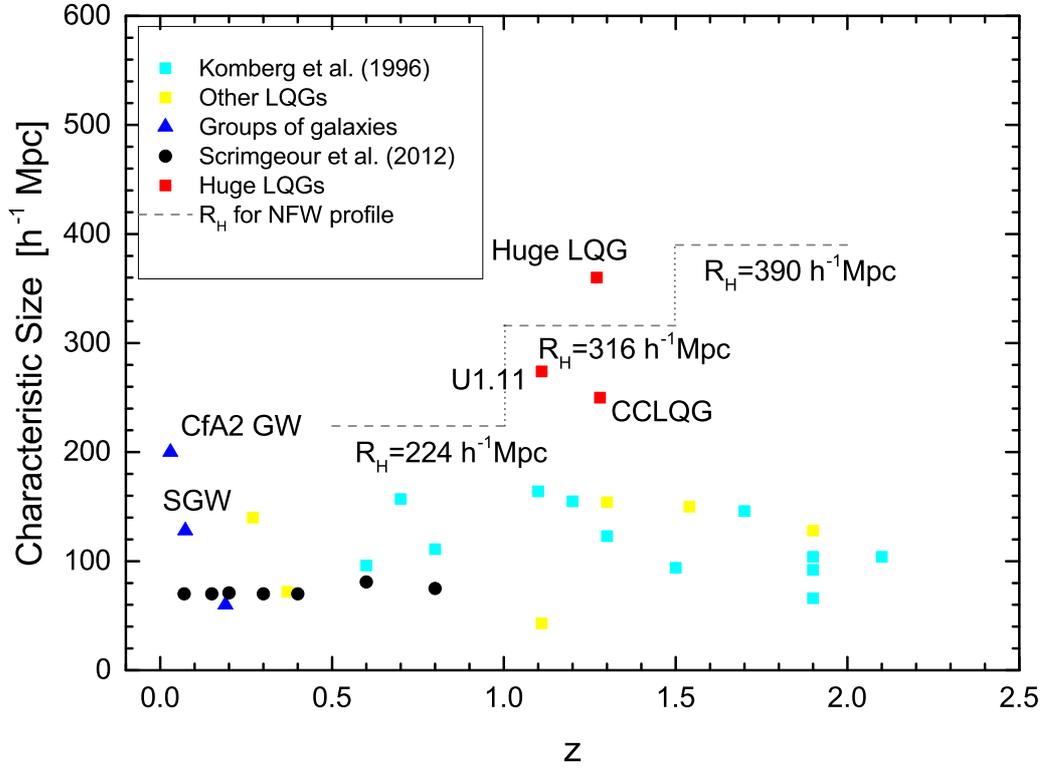}
  \caption{\small{Characteristic size of the LQGs and other large-scale structures. The data are listed in Table \ref{table2} and Table \ref{table3}. LQGs are denoted by rectangles and filament of galaxies are denoted by triangles. The 12 LQG samples from Komberg et al. [24] are shown in cyan rectangles. LQGs identified by other groups but not included in this sample are denoted by yellow rectangles. Black solid circles indicate the homogeneity scales of galaxy distribution given by Scrimgeour et al. [11] at different redshifts. The blue solid triangles denote the characteristic size of the groups of galaxies, i.e. the CfA2 Great Wall [1], Sloan Great Wall [2], and a group of 7 Seyfert galaxies discovered by [57]. Red solid rectangles denote those structures of which the characteristic size is larger than the upper limit $\sim 260$ $h^{-1}$Mpc given by Yadav et al. [6], i.e. the U1.27 (i.e. Huge-LQG) [17], the U1.11 and U1.28 (i.e. CCLQG)  [12]. 
The homogeneity scale obtained from theory for the NFW dark halo profiles are shown in dashed gray lines. For the three redshift bins, the results are $R_H\simeq 224$ $h^{-1}$Mpc for $0.5< z\leq 1$, $R_H\simeq 316$ $h^{-1}$Mpc for $1< z\leq 1.5$, and $R_H\simeq 390$ $h^{-1}$Mpc for $1.5< z\lesssim 2$, as are shown in Table \ref{table4}.}}
  \label{lqgs}
\end{figure*}

\begin{table*} 
\caption{Total halo mass $M_{{\rm DMH}}$ and bias $b$ for the LQG samples from Komberg et al. [24]. Column (1): LQG names. Columns (2): numbers of quasars contained in the group. Column (3): redshifit of LQG. Column (4), (7) and (10): total dark halo masses of the quasars in the LQG, estimated respectively from the isothermal profile (equation (\ref{isodmh})), the NFW profile (equation (\ref{NFWdmh})), and the lensing profile (equation (\ref{lensdmh})). Column (5), (8), and (11): smoothing radius calculated from equation (\ref{radius}). Column (6), (9), and (12): bias factor estimated respectively for the three profiles using equation (\ref{bias1}) given by Mo $\&$ White. The `Total Mean' values are averaged over all the LQGs in the sample, while the `Mean' values refers to that of each redshift bin. The values in bold type are of main concerns in the calculations as well as in conclusions and discussions. }
\begin{tabular}[t]{cccccccccccc}
(1) & (2)& (3) & (4) & (5) & (6) & (7) & (8) & (9) & (10) &(11) & (12)\\
 Group   &   Number of  &  $z$   &  $M_{{\rm DMH\_iso}}$ & $R_{{\rm iso}}$  &  $b(R_{{\rm iso}},z)$ & $M_{{\rm DMH\_NFW}}$  & $R_{{\rm NFW}}$  &  $b(R_{{\rm NFW}},z)$ & $M_{{\rm DMH\_lens}}$  & $R_{{\rm lens}}$  &  $b(R_{{\rm lens}},z)$ \\ 
 &  Quasars  &   & $[10^{14} M_{\odot}]$ & [Mpc]   & & $[10^{14} M_{\odot}]$  & [Mpc]  & & $[10^{14} M_{\odot}]$ & [Mpc]  &   \\
\hline
LQG1 & 12 & 0.6 & 2.40 &  8.02 & 3.97 & 1.48 &  6.83 & 3.23 &  0.41 & 4.46 & 2.01    \\
LQG2 & 12 & 0.8 & 2.40 &  8.02 & 4.92 & 1.48 &  6.83 & 3.97 &  0.41 & 4.46 & 2.43   \\
LQG11 & 11 &  0.7  &  2.20 &  7.79 &  4.26 &  1.36 &  6.63 &  3.46 &   0.38 &   4.33 &  2.15 \\
\hline
Mean & - & 0.7 & 2.33 & 7.94 & {\bf 4.38} & 1.44 & 6.76 & {\bf 3.55} & 0.40 & 4.42 & {\bf 2.20} \\
\hline
\hline
LQG10 & 25 &  1.1  & 5.00 & 10.2 &  9.40 &  3.08 &   8.72 &  7.37 &  0.86 &  5.69 &  4.19   \\
LQG12 & 14 &  1.2  & 2.80 &   8.45 &  7.69  & 1.73 &   7.19 &  6.14  &  0.48 &  4.69  &  3.63  \\
LQG3 & 14 &  1.3  &  2.80 &  8.45 & 8.37 & 1.73 & 7.19  &  6.68 &  0.48 & 4.69 & 3.93    \\
LQG6 & 10 &  1.5  &  2.00 & 7.55 &  8.37  & 1.23 &  6.42 &  6.73 &  0.34  &  4.19  &  4.03    \\
\hline
Mean & - & 1.28 & 3.15 & 8.66 & {\bf 8.46} & 1.94 & 7.38 & {\bf 6.73} & 0.54 & 4.82 & {\bf 3.95} \\
\hline
\hline
LQG5 & 13 &  1.7  & 2.60 &  8.24  &  10.98 & 1.60 &  7.01 &  8.75  &  0.45  &  4.58  & 5.12    \\
LQG4 & 14 &  1.9  &  2.80 &  8.45 & 13.07 & 1.73 & 7.19  &  10.37 &  0.48 & 4.69 & 6.01    \\
LQG7 & 10 &  1.9  &  2.00 & 7.55 &11.12 & 1.23 &  6.42 &  8.91 &  0.34  &  4.19  &  5.28     \\
LQG9 & 18 & 1.9 & 3.60 &  9.18 &  14.81 & 2.22  &  7.81 &  11.67  & 0.62  & 5.10  &  6.65     \\
LQG8 & 12 &  2.1 & 2.40 &  8.02 & 13.79 & 1.48 &  6.83 & 10.99 &  0.41 & 4.46 & 6.41   \\
\hline
Mean & - & 1.9 & 2.68 & 8.29 & {\bf 12.75} & 1.65 & 7.05 & {\bf 10.14} & 0.46 & 4.60 & {\bf 5.90} \\
\hline
\hline
Total Mean & - & 1.39 & 2.75 & 8.33 & {\bf 9.23} & 1.70 & 7.09 & {\bf 7.36}  & 0.47 & 4.63 & {\bf 4.32} \\
\hline
\hline
\end{tabular}\label{table1}
\end{table*}

\begin{table*}
\caption{Summery of values of homogeneity scales $R_H$ presented in Figure \ref{fig4} and \ref{fig5}. Column (1): the redshift bin. Column (2) and (3): values of $R_H$ determined respectively by the criteria ${\mathcal N}(<R_H)=1.01$ and $D_2(<R_H)=2.97$ for the isothermal dark halo profile. Column (4) and (5): values of $R_H$ determined respectively by the criteria ${\mathcal N}(<R_H)=1.01$ and $D_2(<R_H)=2.97$ for the NFW dark halo profile. Column (6) and (7): values of $R_H$ determined respectively by the criteria ${\mathcal N}(<R_H)=1.01$ and $D_2(<R_H)=2.97$ for the lensing dark halo profile. The `Mean' values implies an average over the three redshift bin. The values in bold type are of main concerns in conclusions and discussions. 
}
\begin{tabular}[t]{cccccccc}
(1) & (2)& (3) & (4) & (5) & (6) & (7) \\
Redshift &  isothermal $R_H$ for &  isothermal $R_H$ for&  NFW $R_H$ for&  NFW $R_H$ for&  lensing $R_H$ for&  lensing $R_H$ for \\ 
bin & ${\mathcal N}(<R_H)=1.01$ & $D_2(<R_H)=2.97$ & ${\mathcal N}(<R_H)=1.01$ & $D_2(<R_H)=2.97$ & ${\mathcal N}(<R_H)=1.01$ & $D_2(<R_H)=2.97$\\
& [$h^{-1}$ Mpc] & [$h^{-1}$ Mpc] & [$h^{-1}$ Mpc] & [$h^{-1}$ Mpc] & [$h^{-1}$ Mpc] & [$h^{-1}$ Mpc] \\
\hline
$0.5<z\leq1$ & 235&	250	&208	& ${\bf 224}$	&160	&170 \\
$1<z\leq1.5$  & 325&	340&	290	& ${\bf 316}$&	220&	237 \\
$1.5<z\leq2$ & 405&	440&	360&	${\bf 390}$	&275	&295 \\		
\hline		
Mean & 322	& ${\bf 343}$	& 286        & ${\bf 310}$	& 218 &	${\bf 234}$\\
\hline
\hline
\end{tabular}\label{table4}
\end{table*}

\begin{table*}
\caption{Data of the 12 LQG samples from Komberg et al. [24]. They are shown in Figure \ref{lqgs}. Column (1): LQG names. Column (2): redshift of LQG. Column (3): characteristic size of LQG. The `Total Mean' values are averaged over all the LQGs in the sample, while the `Mean' values refers to that of each redshift bin. The values in bold type are of main concerns in conclusions and discussions. 
}
\begin{tabular}[t]{cccccccccccc}
(1) & (2)& (3)  \\
 Group   &   $z$  &  Characteristic  \\ 
 &    & Size  \\
 & & [$h^{-1}$ Mpc]  \\
\hline
LQG1 & 0.6 & 96 \\
LQG2 & 0.8 & 111   \\
LQG11 & 0.7 &  157   \\
\hline
Mean & {\bf 0.7} & {\bf 121}   \\
\hline
\hline
LQG10 & 1.1 &  164   \\
LQG12 & 1.2 &  155   \\
LQG3 & 1.3 &  123     \\
LQG6 & 1.5 &  94    \\
\hline
Mean & {\bf 1.28} & {\bf 134}  \\
\hline
\hline
LQG5 & 1.7 &  146       \\
LQG4 & 1.9 &  104    \\
LQG7 & 1.9 &  92 \\
LQG9 & 1.9 & 66   \\
LQG8 & 2.1 &  104  \\
\hline
Mean & {\bf 1.9} & {\bf 102}  \\
\hline
\hline
Total Mean & 1.39 & 118  \\
\hline
\hline
\end{tabular}\label{table2}
\end{table*}

\begin{table*}
\caption{Properties of other large-scale structures in the Universe. These structures are not included into the calculations in this paper but are plotted in Figure \ref{lqgs} to compare with the theory-predicted $R_H$. Column (1): names of the structures. `LQG' means that the structure is a large quasar group. Column (2): redshift of the structures. Column (3): characteristic size of the structures. Column (4): notes and references.}
\begin{tabular}[t]{cccccccccccc}
(1) & (2)& (3) & (4) \\
 Name   &   $z$  &  Characteristic Size  &  Notes  \\ 
 &    &  [$h^{-1}$ Mpc]  &     \\
\hline
CfA2 Great Wall & 0.03 & 200  & A filament of galaxies discovered by Geller et al. [1].  \\
Sloan Great Wall (SGW) & 0.073 &  128  & A filament of galaxies discovered by Gott et al. [2].    \\
- & 0.19 & 60 & Discovered by Graham et al. [57], \\
& & & it is a group of 7 Seyfert galaxies\footnote{See the reference Maiolino $\&$ Rieke [58] for the term `Seyfert galaxies'}.    \\
Tesch-Engels LQG & 0.27 & 140 & Discovered by Clowes [59] as the first x-ray selected LQG. \\
Webster LQG & 0.37 &  72  & The first discovered LQG, by Webster [60]. \\
CCH LQG & 1.11 &   43  & Discovered by Crampton et al. [61].   \\
CC-group & 1.3 & 154 &  Discovered by Clowes et al. [4] and [62].\\
Newman LQG (U1.54) & 1.54 &  150 & Discovered by Newman P.R. et al. See the references \\ 
& & & Clowes et al. [5] and [59]. \\
- & 1.9 &  128  &  Discovered by Graham et al. [57].  \\
U1.11 & 1.11 &  274 & A LQG discovered by Clowes et al. [5].\\
CCLQG (U1.28) &1.28 &  250 & Discovered by Clowes et al. [5]. \\
Huge-LQG (U1.27) &1.27 &  360 & Discovered by Clowes et al. [17].\\
\hline
\hline
\end{tabular}\label{table3}
\end{table*}

\end{document}